\documentclass[%
reprint,
amsmath,amssymb,
aps,
pra,
]{revtex4-2}
\usepackage{color}
\usepackage{mathtools}
\usepackage{comment}
\usepackage{float}
\usepackage{graphicx}
\usepackage{dcolumn}
\usepackage{bm}
\usepackage{hyperref}

\begin{document}

\preprint{APS/123-QED}

\title{Interference and Bell States in q-Deformed Quantum Oscillator\\
a Wigner Function Perspective
}

\author{Efe T\"urbedar}
\email{efe.turbedar@ogr.iu.edu.tr}
\author{Ferhat Nutku}
\email{fnutku@istanbul.edu.tr}
\affiliation{Department of Physics, Faculty of Science, \.{I}stanbul University, Vezneciler, \.{I}stanbul, 34134, Turkey}

\date{\today}
\begin{abstract}
In this paper, we investigate the interference and Bell states of a $q$-deformed harmonic oscillator. The Wigner functions of the interference states and the four Bell states are calculated and discussed. It is shown that in the case where $q\rightarrow0$ one can get cat-like states, and in the case where $q\rightarrow1$ one gets the properties of a quantum harmonic oscillator.
\end{abstract}

\maketitle

\section{Introduction}\label{sec:intro}

In the last few years, there has been a growing interest in q-deformed algebras, and the corresponding physical systems. There has been active research topics in q-deformed systems such as in the form of 
quantum Otto engines\cite{Ozaydin2023}, boson algebras\cite{Altintas2021}, and quantum logic gates\cite{Altintas2013}. There are also several works related to exploring superposition of wave functions\cite{Alomeare2024,Jafarov2010}. The superpositions of the stationary states of the q-deformed harmonic oscillator are analyzed with the use of the Wigner function. The Wigner function is a real-valued quasi-probability distribution on phase space, defined via the Weyl transform of a quantum state's density operator, that provides a complete representation of the state's position, and momentum correlations. The Wigner functions of the superposition states have been found to exhibit intriguing properties such as sub-Planck structures\cite{Jafarov2007,Jafarov2010}, and entanglement\cite{Alomeare2024}. Both of these properties have key features to aid in overcoming current challenges in quantum computing. Entanglement generation makes it possible to use $q$-deformed oscillator in the context of quantum computing, and sub-Planck structures can achieve higher quality measurements\cite{Panigrahi2011}.
One of the required conditions for a quantum computer to operate is that the 
system must have an anharmonic energy spectrum which means the energy differences between consequtive levels should not be equal.
The reason for this is that, in a system with a harmonic spectrum, transitions can occur between multiple levels simultaneously, making it difficult to determine which levels are responsible for the transition. \cite{sorensen2024}.
Since the $q$-deformed harmonic oscillator has an anharmonic energy spectrum, 
it can be a candidate for the construction of a quantum computer.
Taking these into account, $q$-deformed oscillator systems starts to look like
a substitute for constructing a new type of quantum computer. Motivated by 
this glimmer of potential for use in quantum computing, we made this paper to
serve as a reference for a future researcher working with qubits that are in a
superposition of two stationary states of the $q$-deformed oscillator. 
Furthermore, we chose to explore the natural first step, akin to writing 
``Hello World!'' using regular bits via by forming Bell states.
    
The article is prepared as follows: in Sec.\,\ref{sec:qho} $q$-deformed 
harmonic oscillator wavefunction is given, in Sec.\,\ref{sec:interference}
simplified analytical expression for Wigner quasi-probability distribution
function is presented. The importance and form of Bell states are given
very briefly in Sec.\,\ref{sec:Bell}. Afterwards, in Sec.\,\ref{sec:Bellqho}
Wigner functions for four Bell states are constructed and corresponding
phase space plots are discussed in Sec.\,\ref{sec:discussion}, final
remarks are given in conclusion Sec.\,\ref{sec:concs}.

\section{$Q$-Deformed Harmonic Oscillator}\label{sec:qho}

The stationary states of the \emph{q}-deformed quantum harmonic oscillator in the \(x\)-representation are defined as \cite{Jafarov2010},
\begin{equation}
\label{eq:wavefunction}
\psi_n^{q\mathrm{HO}}(x) = c_n 
\sum_{k=0}^{n} \frac{\bigl(q^{-n}; q\bigr)_{k}}{\bigl(q; q\bigr)_{k}} \,
q^{n k - \tfrac{k^2}{2}} \,
e^{-2 i \lambda h x k} 
\,e^{-\lambda x^{2}}
\end{equation}
where $\lambda$ parameter is 
\begin{equation}
\label{eq:lambda}
\lambda \;=\; \frac{m \,\omega}{2\,\hbar}
\end{equation}
\(c_{n}\) is the normalization constant,
\begin{equation}
\label{eq:cn}
c_{n} \;=\; 
\left(\frac{2\,\lambda}{\pi}\right)^{\!\tfrac{1}{4}}\,
i^n\,
q^{\tfrac{n}{2}}\,
\bigl(q;q\bigr)_{n}^{-\tfrac{1}{2}}
\end{equation}
and \(h\) is the deformation parameter which is related to $q$
as the following,
\begin{equation}
\label{eq:q}
q \;=\; e^{-\lambda\,h^{2}}, 
\quad 0 < q < 1,
\quad 0 < h < +\infty.
\end{equation}
Inside the summation of Eq.~\ref{eq:wavefunction}, \(\bigl(q;q\bigr)_{k}\) denotes the \emph{$q$-Pochhammer symbol} \cite{GasperRahman2004, koekoek1996}, defined by 
\begin{equation}
\label{eq:q-shifted-factorial}
(a;q)_0 = 1, 
\quad
(a;q)_k 
= \prod_{n=0}^{k-1}\bigl(1-a\,q^n\bigr).
\end{equation}
It is known that in the limit $q$ $\rightarrow$ 1, a wave function for $q$-deformed quantum harmonic oscillator becomes a stationary state wave function of an ordinary quantum mechanical harmonic oscillator\cite{Jafarov2007}.

\section{Interference States for $Q$-Deformed Harmonic Oscillator}\label{sec:interference}

In the work of Alomeare et al.\cite{Alomeare2024}, Wigner 
quasi-probability distribution function beloging to some superpositions of $q$-oscillator pure states were shown. 
Superposition of two pure states with proper probability
amplitudes yield interference patterns similar to cat states, and 
four pure states yield sub-Planck structures. Cat states are a powerful tool in quantum computing. They are invaluable for quantum error correction, entanglement generation, and certain computational protocols. These aspects motivated us to further investigate the $q$-deformed harmonic oscillator for the researchers studying on	quantum computing.

A particular superposition of the $q$-deformed oscillator 
which consists of two pure states is 
$\psi_{nm}=a\psi_n^{q\mathrm{HO}}(x) + b\psi_m^{q\mathrm{HO}}(x)$.
Wigner function for a single particle wave function \cite{Wigner1932, Agarwal2004} is defined as 
\begin{equation}
W(x,p)=\frac{-1}{2\pi\hbar}\int_{-\infty}^{\infty}e^{-ipy/\hbar}\
\psi\left(x+\frac{y}{2}\right) \psi^{*}\left(x-\frac{y}{2}\right)dy.
\end{equation}
This formula can be applied to find a general Wigner function 
expression for superposition of two $q$-deformed harmonic 
oscillator states having different deformation parameters $q_A$ and $q_B$ 
with quantum numbers $n$ and $m$, respectively.
After some calculation and using Gauss integral formula,
\begin{equation}
\int_{-\infty}^{\infty}e^{-\left(a_2 x^2+a_1 x+a_0\right)}\,
dx = \
\sqrt{\frac{\pi}{a_2}} e^{\frac{a_ 1^2}{4 a_ 2}-a_ 0} 
\end{equation}
where $a_2>0 \in \mathbb{R}$ and $a_0, a_1 \in \mathbb{C}$, one
can obtain the following formula,
\begin{equation}
\begin{aligned}
W_{n,m}(x&, p)=\frac{-1}{2\pi\hbar}\sqrt{\frac{2\pi}{\lambda}}e^{-2\lambda x^2}\\
&\times \Bigl[\, |a|^2\,W_{n,q_a,n,q_a}(x,p) + a^*b\,W_{n,q_a,m,q_b}(x,p)\\
& + b^*a\,W_{m,q_b,n,q_a}(x,p) + |b|^2\,W_{m,q_b,m,q_b}(x,p)\, \Bigr]
\end{aligned}
\label{eq:bellwigner}
\end{equation}
where $W_{n,n}, W_{n,m}, W_{m,n}, W_{m,m}$ terms can be 
obtained from a generic $W_{j,l}$ function which is defined as
the following
\begin{equation}
\begin{aligned}
W_{j,l}(x, p) &= c_i^{*}c_j \\
&\times \sum_{k=0}^{j} \sum_{s=0}^{l}{\mathbb{B}_{q_a,q_b}^{j,l}(k,s)}\\
&\times e^{2ix\lambda(h_a k - h_b s)}\\
&\times e^{\frac{-\left(\lambda h_a k+ \lambda h_b s+
\frac{p}{\hbar}\right)^2}{2\lambda}}
\end{aligned}
\label{eq:Psi}
\end{equation}
where $\mathbb{B}_{q_a,q_b}^{j,l}(k, s)$ is defined as
\begin{equation}
    \begin{aligned}\mathbb{B}_{q_a,q_b}^{j,l}(k, s)=\frac{ \left(q_a^{-j};q_a\right)_k \left(q_b^{-l};q_b\right)_s}{(q_a;q_a)_k (q_b;q_b)_s}q_a^{jk-\frac{k^2}{2}}q_b^{ls-\frac{s^2}{2}}
        \end{aligned}
    \label{eq:binomial}
\end{equation}

\begin{figure}[htbp]
\centering
\includegraphics[width=0.9 \linewidth]{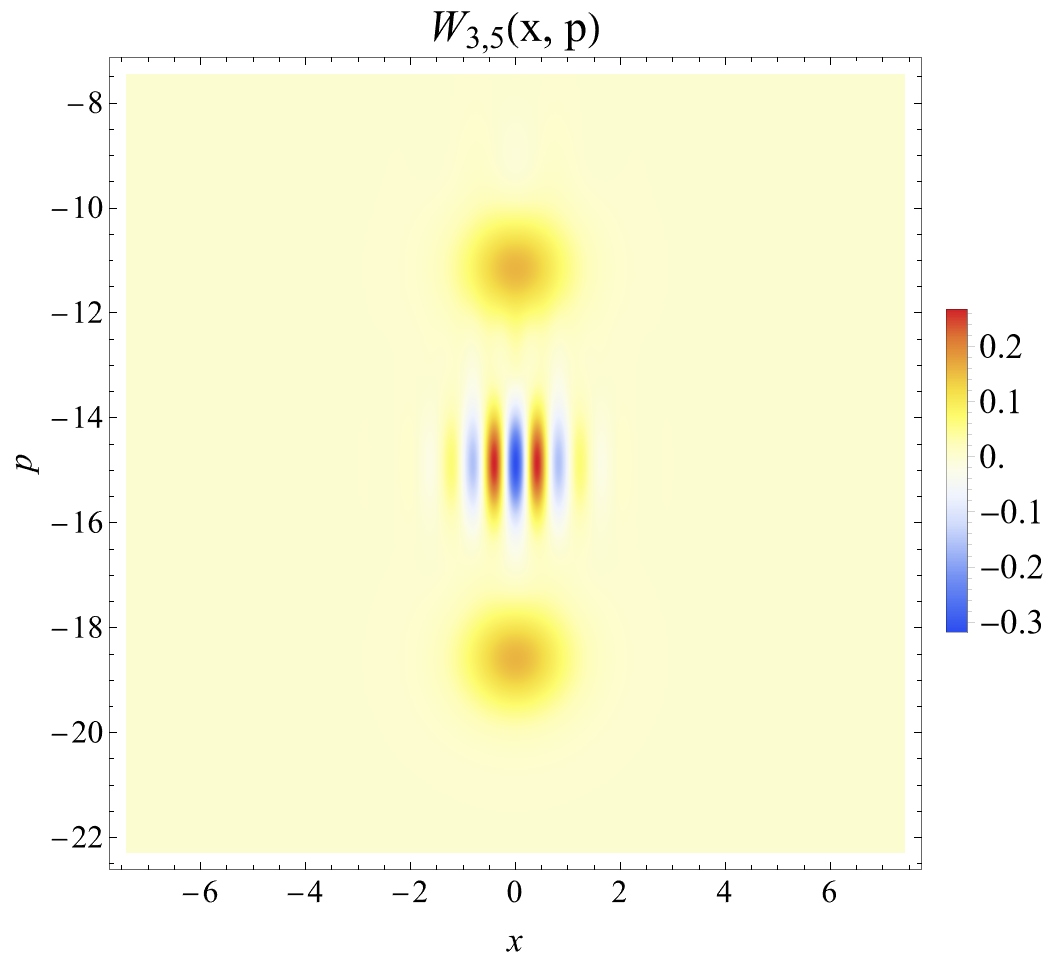}
\caption{Wigner quasi-probability distribution of 
superposition of two $q$-deformed harmonic oscillator pure states. Selected parameters are $n=3, m=5, q=0.001$ and $a=b=1/\sqrt{2}$.}
\label{fig:interference3_5}
\end{figure}

In Fig.~\ref{fig:interference3_5} Wigner quasi-probability distribution
for the superposition of $2^{nd}$ and $3^{rd}$ states with the
same probability coeffcient $1/\sqrt{2}$ are given.
In the aforementioned work \cite{Alomeare2024}, this superposition is shown to have no entanglement present for a fixed deformation parameter 
$q_a=q_b=q$ case.

\section{Bell States}\label{sec:Bell}

Entanglement is without a doubt one of the most interesting phenomena in quantum 
mechanics, and we utilize it many ways, almost exclusively in quantum computation.
Entanglement is often generated as a resource to be used on applications such as
quantum teleportation and superdense coding. The most basic example of entanglement
generation can be formed in terms of Bell states. Bell states are a set of four 
maximally entangled two-particle states that form an orthonormal basis for tensor product states living in the Hilbert space $\mathbb{C}^2 \otimes \mathbb{C}^2$.
These states are crucial in entanglement generation.

The Bell states are defined as the following,
\begin{subequations}\label{eq:bell_states}
    \begin{align}
        |\Psi^+\rangle &= \frac{\,|0\rangle_A \otimes |1\rangle_B + |1\rangle_A \otimes |0\rangle_B\,}{\sqrt{2}} \label{eq:bell_a} \\
        |\Psi^-\rangle &= \frac{\,|0\rangle_A \otimes |1\rangle_B - |1\rangle_A \otimes |0\rangle_B\,}{\sqrt{2}} \label{eq:bell_b} \\
        |\Phi^+\rangle &= \frac{\,|0\rangle_A \otimes |0\rangle_B + |1\rangle_A \otimes |1\rangle_B\,}{\sqrt{2}} \label{eq:bell_c} \\
        |\Phi^-\rangle &= \frac{\,|0\rangle_A \otimes |0\rangle_B - |1\rangle_A \otimes |1\rangle_B\,}{\sqrt{2}} \label{eq:bell_d}
    \end{align}
\end{subequations}
where the indices $A$ and $B$ denote the spatially separated particles.

\section{Bell States for q-Deformed Harmonic Oscillator}\label{sec:Bellqho}

For our purposes, the two-state particles are selected as superpositions of $q$-deformed harmonic oscillators in the $n^{th}$ and $m^{th}$ states.
The Wigner function of a two-particle system described by a wave function $\psi$ is defined as \cite{Bhatt2008},
\begin{equation}
        \begin{aligned}
        W(x_A, p_A, x_B, p_B) &= \frac{1}{4 \pi^2 \hbar^2} \int_{-\infty}^{\infty} \int_{-\infty}^{\infty} e^{i\left(\frac{p_A y_A + p_B y_B}{\hbar}\right)}\\
        & \quad \times \psi\left(x_A - \frac{y_A}{2},\, x_B - \frac{y_B}{2}\right) \\
       & \quad \times \psi^*\left(x_A + \frac{y_A}{2},\, x_B + \frac{y_B}{2}\right)dy_Ady_B
    \end{aligned}
\end{equation}
This definition of the Wigner function uses the wave function instead of a ket, so we need to use the $x$-space representation of the Bell states. This is simply obtained by taking the inner product $\langle x_A, x_B | \beta \rangle$ where $\beta$ is an arbitrary Bell state. For example for $\Psi^+$ we get,
\begin{equation}
\langle x_A, x_B | \Psi^+ \rangle = \frac{1}{\sqrt{2}} \left( \psi_0(x_A) \, \psi_1(x_B) + \psi_1(x_A) \, \psi_0(x_B) \right).
\end{equation}
Of course in the above equation we will have $\psi_0 $ and $\psi_1 $ 
corresponding to $\psi_n$ and $\psi_m$, respectively.

Calculating the Wigner function of the Bell states using the $n^{th}$ and $m^{th}$ order $q$-deformed harmonic oscillator wave functions we obtain 4 double integrals.
In the calculation, first and last terms create the Gaussians
that correspond to the $n^{th}$ and $m^{th}$ states in the Wigner phase space, and the second and third terms combine to create a single term that is responsible from the interference. This becomes more apparent when the phase spaces are plotted with by 
choosing high deformation values.
After a long calculation the Wigner function of the Bell states for the 
$q$-deformed harmonic oscillators system can be expressed explicitly as,
\begin{equation}
  \begin{aligned}
    W_{n,m}(x_A&, x_B, p_A, p_B)_{\Psi^{\pm}|\Phi^{\pm}}=\frac{e^{-2 \lambda_A x_A^2 - 2 \lambda_B x_B^2}}{4 \pi \hbar^2 \sqrt{\lambda_A \lambda_B}}\\
    &\quad \times \Bigl[\, W_1(x_A, x_B, p_A, p_B)_{\Psi_1|\Phi_1}\\ 
    &\quad \pm  W_2(x_A, x_B, p_A, p_B)_{\Psi_2|\Phi_2}\\ 
    &\quad  W_3(x_A, x_B, p_A, p_B)_{\Psi_3|\Phi_3} \, \Bigr]
  \end{aligned}
\label{eq:bellwigner}
\end{equation}
Here are the $W_{\Psi|\Phi}$ terms are
\begin{equation}
\begin{aligned}
W_1(x_A,&x_B,p_A,p_B)_{\Psi_1} =\; |c_{n,A}|^2\,|c_{m,B}|^2 \\& \times
\sum_{k_1,k_2=0}^{n} \sum_{s_1,s_2=0}^{m}{\mathbb{B}_{q_A}^{n, n}(k_1, k_2)}
    {\mathbb{B}_{q_B}^{m, m}(s_1, s_2)} \\
 &\times \epsilon(k_1 ,k_2,s_1 ,s_2)\,\kappa(k_1 ,k_2,s_1 ,s_2)\,\\\\
W_2(x_A,&x_B,p_A,p_B)_{\Psi_2} =\;  2\,c_{m,A}^*\,c_{n,A}\,c_{n,B}^*\,c_{m,B} \\& \times\sum_{k_1,k_2=0}^{n} \sum_{s_1,s_2=0}^{m} 
 {\mathbb{B}_{q_A}^{n, m}(k_1, s_2)}
         {\mathbb{B}_{q_B}^{m, n}(s_1, k_2)} \\
&\times  \epsilon(k_1 ,s_2,s_1 ,k_2)\,\kappa(k_1 ,s_2,s_1 ,k_2)
\\\\
W_3(x_A,&x_B,p_A,p_B)_{\Psi_3} =\;  |c_{m,A}|^2\,|c_{n,B}|^2 \\&\times\sum_{k_1,k_2=0}^{n}  \sum_{s_1,s_2=0}^{m}  {\mathbb{B}_{q_A}^{m, m}(s_1, s_2)}
         {\mathbb{B}_{q_B}^{n, n}(k_1, k_2)} \\
& \times \epsilon(s_1 ,s_2,k_1 ,k_2)\,\kappa(s_1 ,s_2,k_1 ,k_2)
\end{aligned}
\label{eq:Psi}
\end{equation}
and
\begin{equation}
\begin{aligned}
W_1(x_A,&x_B,p_A,p_B) _{\Phi_1}=\;  |c_{n,A}|^2\,|c_{n,B}|^2  \\&\times\sum_{k_1,k_2,s_1,s_2=0}^{n} \Biggl( {\mathbb{B}_{q_A}^{n, n}( k_1, k_2)}{\mathbb{B}_{q_B}^{n, n}( s_1, s_2)} \\
&  \times \epsilon(k_1 ,k_2,s_1 ,s_2)\,\kappa(k_1 ,k_2,s_1 ,s_2)
\Biggr)\\\\
W_2(x_A,&x_B,p_A,p_B)_{\Phi_2} =\; \,2\,c^{*}_{n,A}\,c_{m,A}\,c^{*}_{n,B}\,c_{m,B} \\
&\times \sum_{k_1,k_2=0}^{n} \sum_{s_1,s_2=0}^{m} \biggl( {\mathbb{B}_{q_A}^{n, m}( k_1, s_1)}{\mathbb{B}_{q_B}^{n, m}( k_2, s_2)} \\
 &\times \epsilon(k_1 ,s_1,k_2 ,s_2)\,\kappa(k_1 ,s_1,k_2 ,s_2)
\biggr)\\\\
W_3(x_A,&x_B,p_A,p_B) _{\Phi_3}=\; |c_{m,A}|^2\,|c_{m,B}|^2 \\&\times\sum_{k_1,k_2,s_1,s_2=0}^{m} \biggl( {\mathbb{B}_{q_A}^{m, m}( k_1, k_2)}{\mathbb{B}_{q_B}^{m, m}( s_1, s_2)}\\
  &\times \epsilon(s_1 ,s_2,k_1 ,k_2)\,\kappa(s_1 ,s_2,k_1 ,k_2)
\biggr)
\end{aligned}
\label{eq:Phi3}
\end{equation}
where each $\mathbb{B}_q$-factor is given as
\begin{equation}
    \begin{aligned}\mathbb{B}_{q}^{n, m}( k, s)=\frac{ \left(q^{-n};q\right){}_k \left(q^{-m};q\right){}_s}{(q;q)_k (q;q)_s}q^{k n-\frac{k^2}{2}+s m-\frac{s^2}{2}}
        \end{aligned}
    \label{eq:binomial}
\end{equation}
and $\kappa$ and $\epsilon$ are defined as
\begin{equation}
    \begin{aligned}
        \kappa(k_1 ,s_1,k_2 ,s_2)&= \cos\biggr(2 \lambda_A x_A h_A (k_1 - k_2) 
        \\  &\quad \quad \quad+2 \lambda_B x_B h_B (s_1 - s_2) \biggl),
        \\
        \epsilon(k_1 ,s_1,k_2 ,s_2) &= e^{\left(\frac{\left((k_1 + k_2) h_A \lambda_A + \frac{p_A}{\hbar}\right)^2}{-2 \lambda_A}\right)} \\ 
        &\,\,\,\times e^{\left( \frac{\left((s_1 + s_2) h_B \lambda_B + \frac{p_B}{\hbar}\right)^2}{-2 \lambda_B}\right)}.
    \end{aligned}
    \label{eq:epsilon}
\end{equation}
Here $W_1$ and $W_3$  are the Wigner functions of the $n^{th}$ and $m^{th}$ states while $W_2$ is the function of the  interference.
Here the $x$-dependence of $\kappa$ and  $p$-dependence of $\epsilon$ functions are omitted in the notation for the sake of brevity. One more thing to take note is that the similarity between $\mathbb{B}_{q}^{n, m}$ and the $_3\phi_2$ basic hypergeometric function. They look similar and other works have found that the terms inside the Wigner function of some $q$-deformed systems can be expressed in terms of this hypergeometric functions\cite{Jafarov2010}. However we have not been able to reduce our expressions using the identities and well known functions from the theory of 
$q$-series and basic hypergeometric series.

\section{Discussions}\label{sec:discussion}

\begin{figure}[!htp]
\centering
\includegraphics[width=0.9 \linewidth]{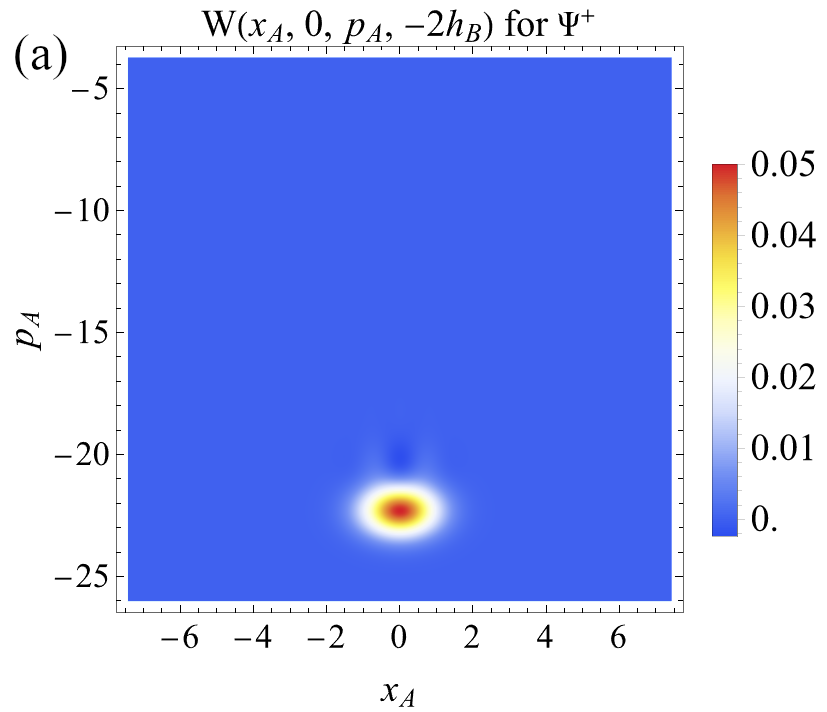} \\
\includegraphics[width=0.9 \linewidth]{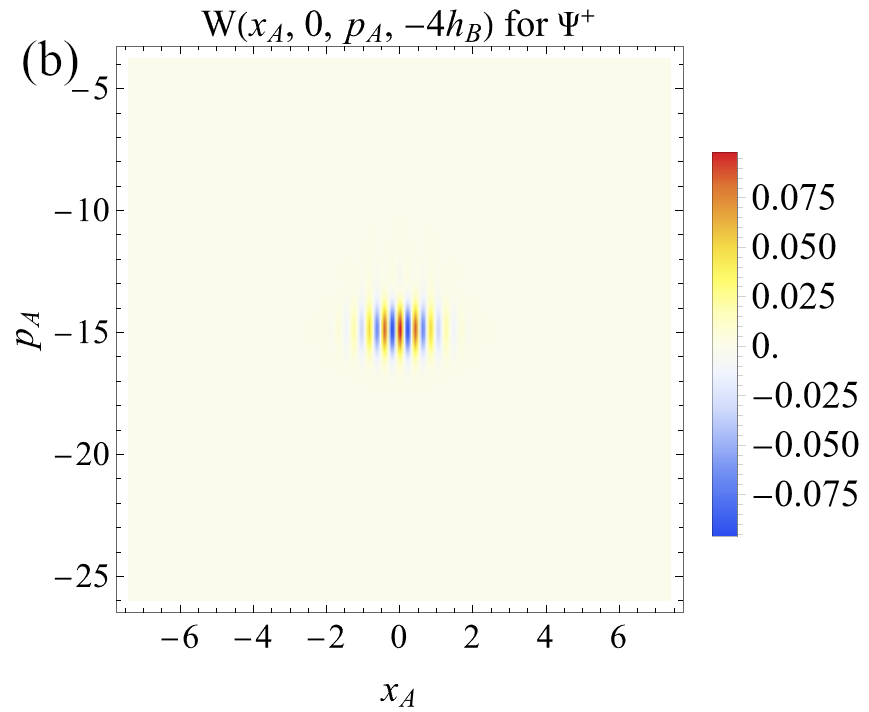} \\
\includegraphics[width=0.9 \linewidth]{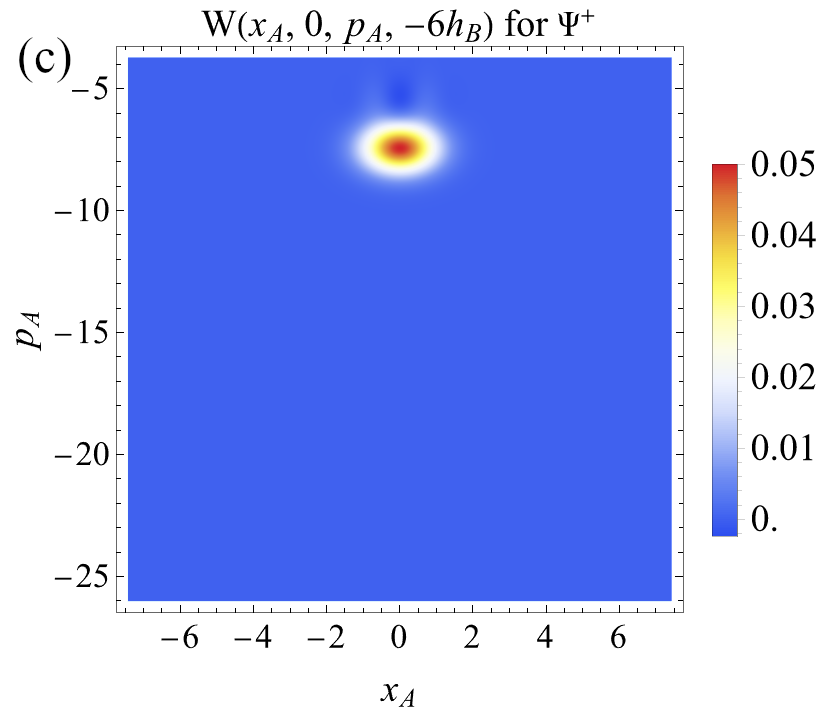} 
\caption{Wigner quasi-probability distribution of $\Psi^+$ Bell 
state where the position and momentum of particle $B$ are fixed.
Selected parameters are $n=2, m=6$ and $q_A=q_B=0.001$.}
\label{fig:psipxApA}
\end{figure}

\begin{figure}[!htp]
\centering
\includegraphics[width=0.9 \linewidth]{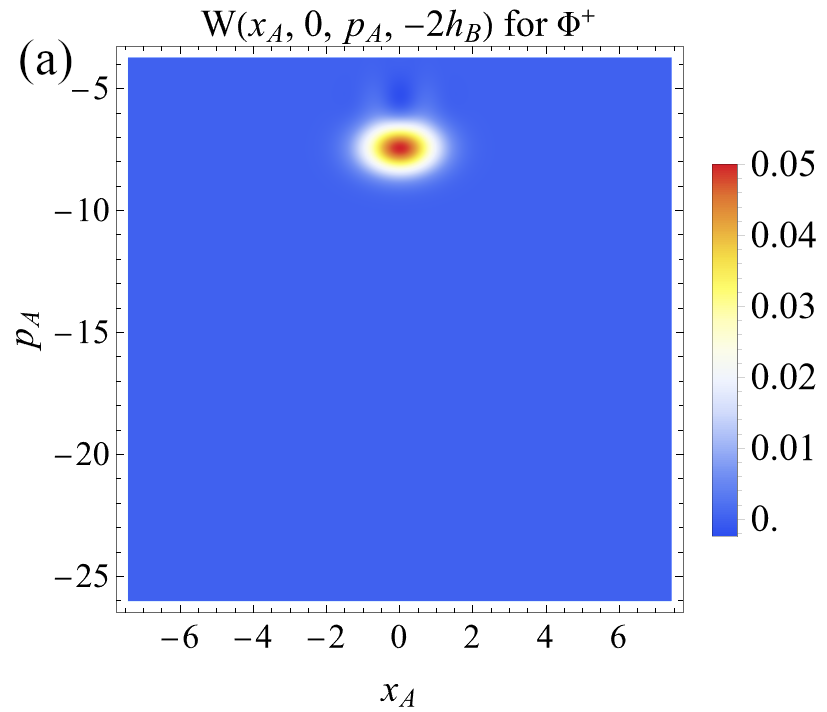} \\
\includegraphics[width=0.9 \linewidth]{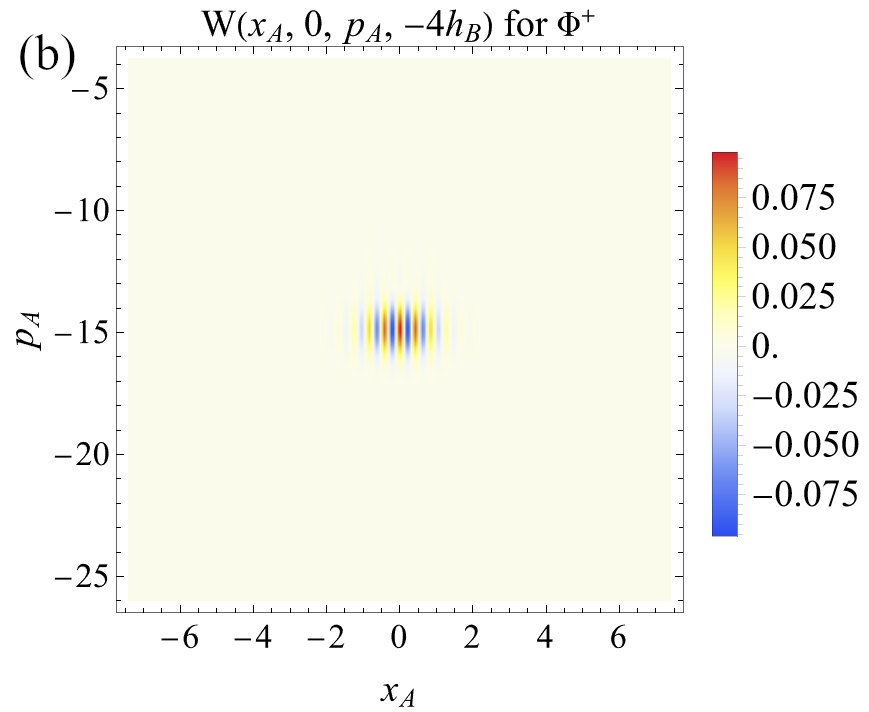} \\
\includegraphics[width=0.9 \linewidth]{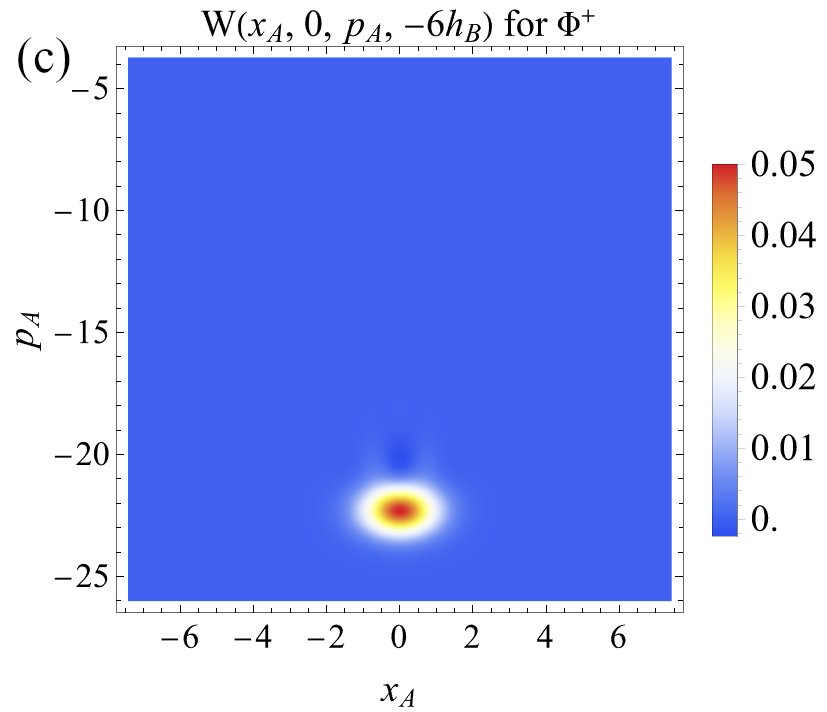} 
\caption{Wigner quasi-probability distribution of $\Phi^+$ Bell 
state where the position and momentum of particle $B$ are fixed.
Selected parameters are $n=2, m=6$ and $q_A=q_B=0.001$.}
\label{fig:phipxApA}
\end{figure}

Since the Wigner function of a system contains all of the information about
the system, we can visually observe different properties of Bell states and 
the $q$-deformed quantum harmonic oscillator by plotting out ``slices'' of 
the 4-dimensional Wigner functions. Let us first look at the $\Psi^+$ and $
\Phi^+$ Bell states where $n=2$,  $m=6$ and we have a large deformation in 
both particles: $q_A=q_B=0.001$. Additionally, we take $\omega= m=\hbar=1$ for simplicity. To plot these 4-dimensional functions we 
will take 2-dimensional slices by fixing 2 paramaters, and because they are
simply more interesting, we will mostly look at the plots where we fix,
space and momentum parameters of one particle, or space parameters of both 
particles. These alone give a good 
enough idea of how the four-dimensional Wigner functions behave.
In Fig.\,\ref{fig:psipxApA}a for the $\Psi^+$ state, we can see that when we select
particle B to be in $x_B =0$ and have momentum $p_B = -2h_B$ (for $q=0.001, 
h\approx3.716$), we see a Gaussian located around $x_A=0,p_A=-6h_A$.
And in  Fig.\, \ref{fig:psipxApA}c when we select $x_B =0$ and  $p_B = -6h_B$,  
we see that the Gaussian
describing the particle $A$ is now located around $x_A=0,p_A=-2h_A$. Therefore, when we
chose particle B to be in $x_B =0$ and to have momentum $p_B = -2h_B$, we essentially
measured particle B to be in the state $n=2$.  And from the definition of $\Psi^+$ Bell 
state it follows that if one particle is measured to be in one state, the other particle 
must be in the complementary state, which is the $m^{th}$ state. Similarly in Fig.\,
\ref{fig:phipxApA}a and Fig.\,\ref{fig:phipxApA}c with $\Phi^+$, when the state of one 
particle is measured, the other is in the same state. The locations of these Gaussians – in other words, the localized states – depend on the $h$ parameter.
It was previously shown that the Wigner 
functions of the $q$-deformed harmonic oscillator had a displacement towards negative 
momentum values depending on the $q$ parameter\cite{Jafarov2007}. This displacement 
results in the $n^{th}$ state being located around $p_B = -n h_B$ and the $m^{th}$ state 
$p_B = -mh_B$.   
But when we look at the slice where $p_B=-\frac{n+m}{2}h_B$ as seen from 
Fig.\,\ref{fig:psipxApA}b and Fig.\,\ref{fig:phipxApA}b, 
we see one more thing, which is an interference pattern that
arises from the quantum coherence between the two distinct states. 
If the system were merely a classical mixture, you would not see this interference. 
The fringes indicate that the system is genuinely in a superposition, not in a definite 
state, until a measurement is made. 

At first glance, $\Psi^+$ and $\Phi^+$ seem to have similar interference patterns as
seen from Fig.\,\ref{fig:psipxApA}b and Fig.\,\ref{fig:phipxApA}b.
However, fixing the momentum parameters reveals that their 
interference fringes are rotated by $90^\circ$ relative to each other, i.e. 
the spatial density distribution for $\Psi^+$ is obtained from that of 
$\Phi^+$ by the transformation $(x,y) \to (-y,x)$, which reflects the underlying local unitary Pauli rotation which connects these states. Moreover, unsurprisingly, the $\Psi^-$ and $\Phi^-$ states are negatives of their + counterparts. This is apparent from the fact that the only difference between the + and - states is that the middle $W_2$ term in both formulas—which is actually the interference term and can reveal the interference even without the other terms—is negative rather than positive.
\begin{figure*}[!htp]
\centering
\includegraphics[width=0.45\linewidth]{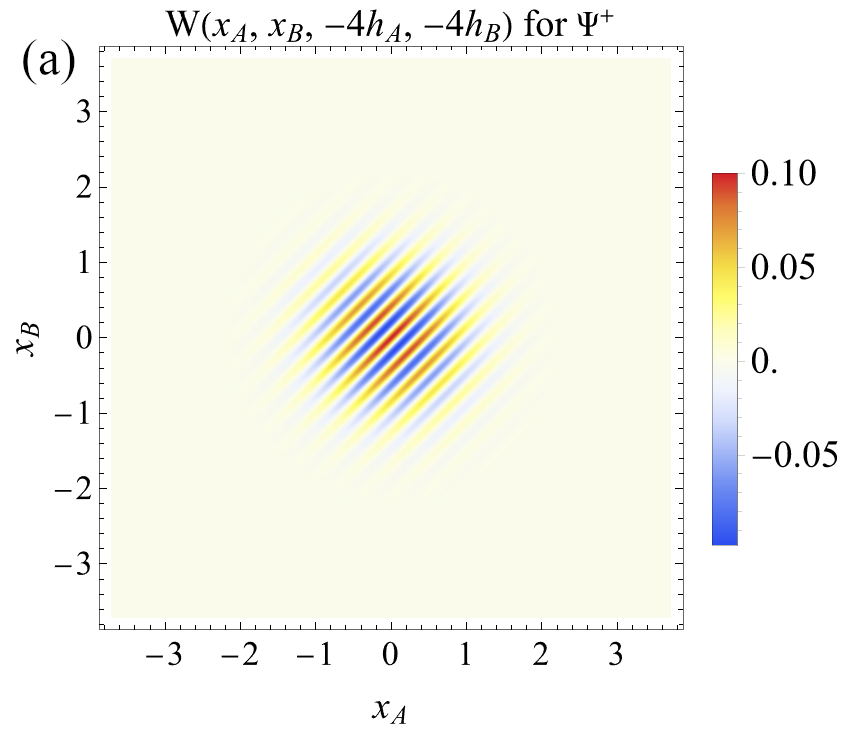}
\includegraphics[width=0.45\linewidth]{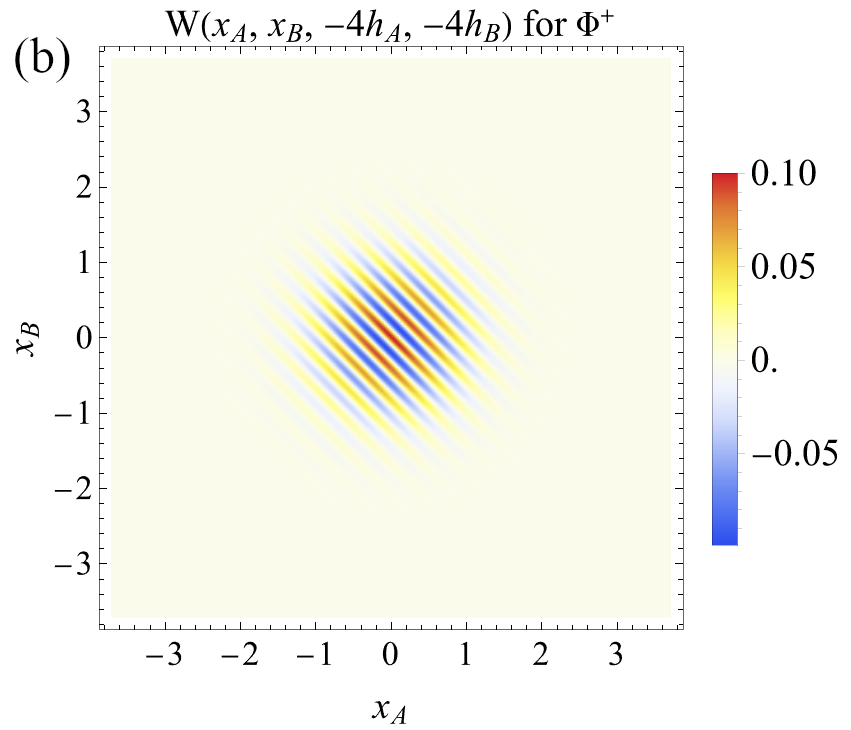}\\
\includegraphics[width=0.45\linewidth]{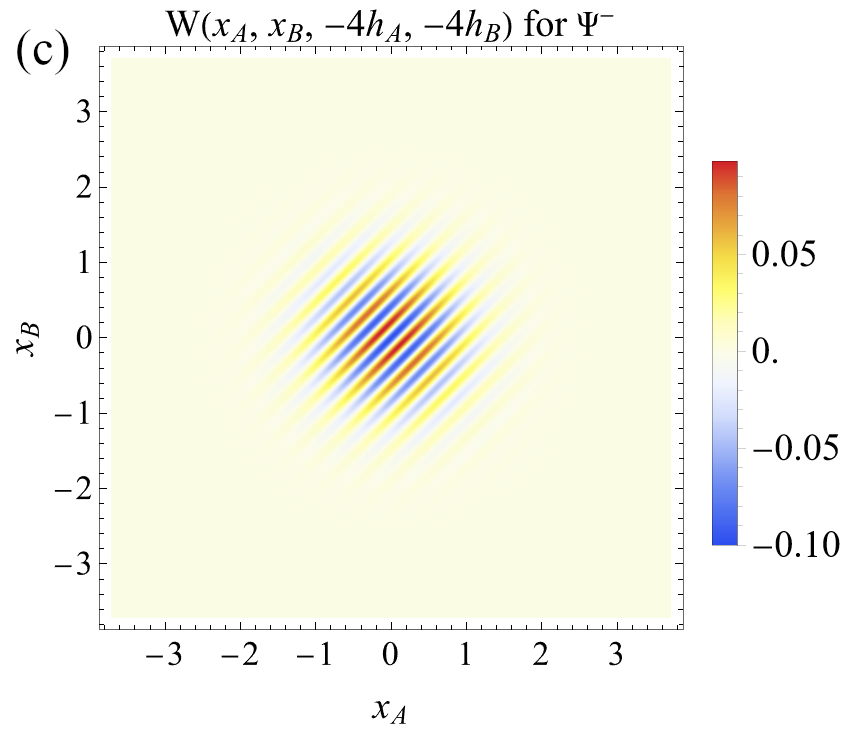}
\includegraphics[width=0.45\linewidth]{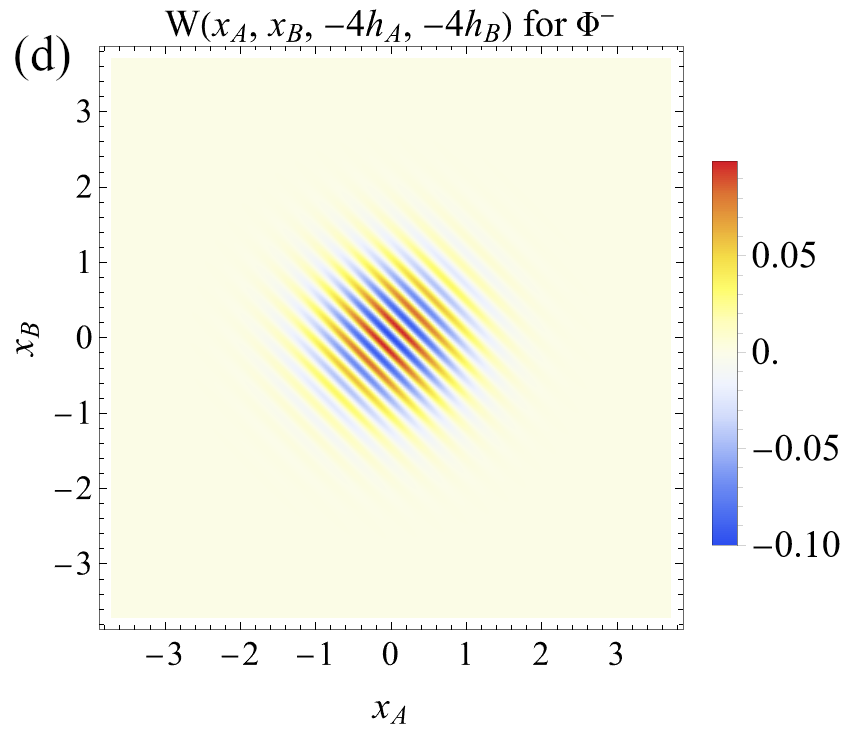}
\caption{Interference patterns in the spatial components 
of the four-dimensional Wigner quasi-probability distribution
of the four Bell States.}
\label{fig:xAxB}
\end{figure*}

\begin{figure}[!htp]
\centering
\includegraphics[width=0.9\linewidth]{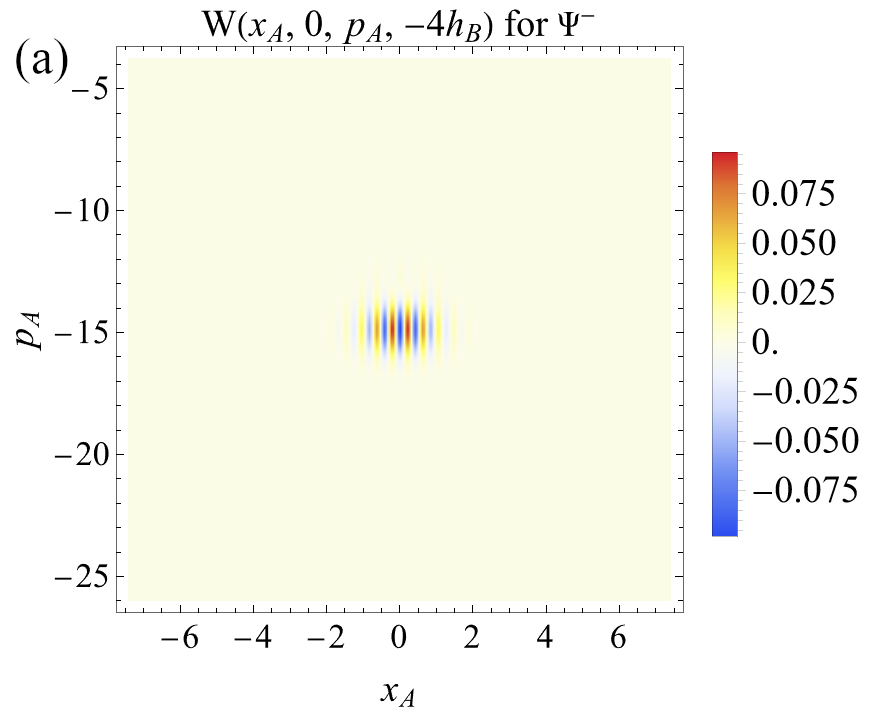} \\
\includegraphics[width=0.9\linewidth]{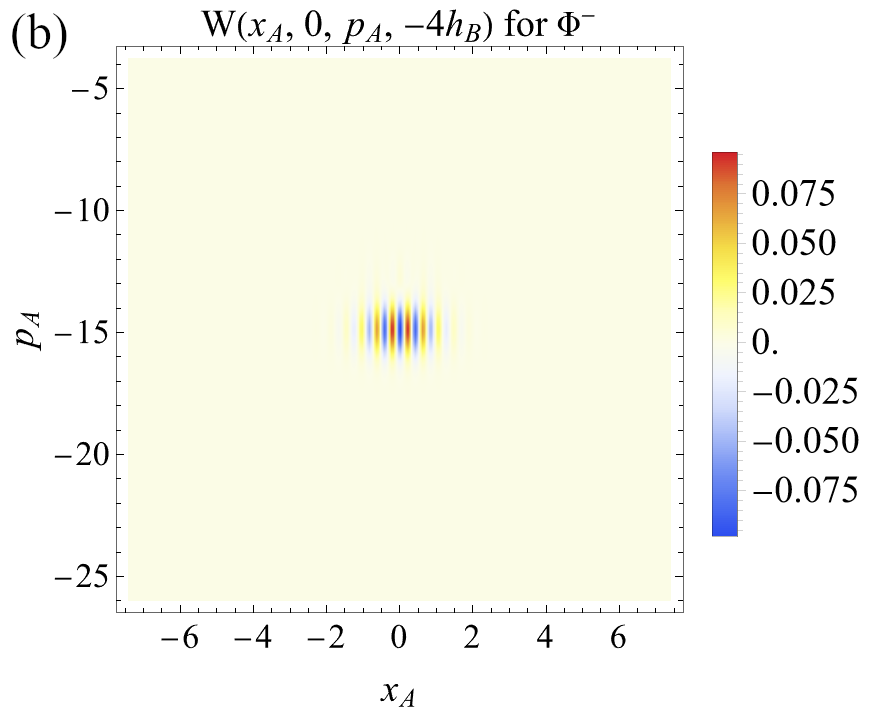} 
\caption{Interference patterns in Wigner quasi-probability distribution  of the asymmetric Bell States, $\Psi^-$ (a) and $\Phi^-$ (b).
Selected parameters are $n=2, m=6$ and $q=0.001$.}
\label{fig:psim_phim}
\end{figure}

Another view of the interference patterns of the $\Psi^-$ and $\Phi^-$ states can be obtained by plotting $p_A$ against $x_A$  where $x_B=0$ $p_B=-\frac{n+m}{2}h_B$. Therefore, once again they are 
the negated versions of their + counterparts, as seen in Fig.\,\ref{fig:psim_phim}. 

\clearpage

\section{Conclusions}\label{sec:concs}

From this research alone, it may not be immediately apparent that the q-deformed 
oscillator presents an advantage. However, possibilities such as generating sub-Planck structures or employing emerging measurement techniques make further investigations of $q$-deformed oscillator systems worthwhile. For example, since each 
stationary state of the $q$-oscillator with large deformation has a corresponding
displacement of the quasi-probability distribution function peak 
towards negative values of the momentum, the $q$ and $\lambda$ parameters 
can perhaps provide an opportunity to fine-tune our system. 
We hope this work is going to lay a small stepping stone for those 
who will follow along the path to making quantum computers out of $q$-deformed harmonic oscillator systems.

\bibliography{references}
\end{document}